\documentclass[12pt]{article}

\usepackage{latexsym}

\usepackage{graphicx}

\begin{document}

\title{Interior perfect fluid scalar-tensor solution}

\author{
     Stoytcho S. Yazadjiev \thanks{E-mail: yazad@phys.uni-sofia.bg}\\
{\footnotesize  Department of Theoretical Physics,
                Faculty of Physics, Sofia University,}\\
{\footnotesize  5 James Bourchier Boulevard, Sofia~1164, Bulgaria }\\
}

\date{}

\maketitle

\begin{abstract}
We present a new exact perfect fluid interior solution for a particular scalar-tensor 
theory. The solution is regular everywhere and has a well defined boundary where the fluid
pressure vanishes. The metric and the dilaton field  match continuously the external solution.   
\end{abstract}


\sloppy

Exact solutions provide a route to better and more deep understanding of 
the inherent nonlinear character of gravity. One of the most difficult tasks is the construction of interior perfect fluid solutions which are of great astrophysical interest. There are some known exact perfect fluid interior solutions in general relativity \cite{KSHMC}. Most of them, however, are  not physically acceptable: the perfect fluid satisfies unrealistic equations of state, the natural energy conditions are violated, the space-time is singular or the solutions have not a well defined boundary. Nevertheless, those solutions are useful since they provide some non-perturbative insight into  the highly nonlinear gravitational phenomena. On the other hand, the exact solutions, even unrealistic, could serve as tests for checking the computer codes which is important for the advent of numerical relativity.                

The most realistic exact interior general relativistic solution is the Schwarzschild interior solution describing a static, spherically symmetric perfect fluid star with a uniform density. This solution has well defined boundary where the fluid pressure vanishes and  matches continuously the external Schwarzschild solution on that boundary. The interior  Schwarzschild solution qualitatively describes  the general case of a static,  spherically symmetric perfect fluid star in general relativity and in particular, it predicts the existence of an upper limit for the ratio of the star mass to the star radius,  known in the general case as the Buchdahl inequality \cite{B}.

General relativity is not the only viable theory of gravity. The most natural alternatives 
of general relativity are the scalar-tensor theories of gravity \cite{DEF},\cite{Will}. In these theories gravity is mediated not only by the metric of space-time but also by a scalar field. 
The importance of scalar-tensor theories is related to the unifying theories, such as string theory and higher dimensions gravity theories, which in their low energy limit predict 
the existence of a scalar partner of the tensor graviton -the dilaton.

Scalar-tensor theories contain arbitrary functions of the dilaton field determining 
the coupling between the dilaton and the matter fields. With a proper choice of the coupling functions, scalar-tensor theories can pass all experimental constraints. In the weak field regime the predictions of the scalar-tensor theories are very close to those of general relativity. However, in the strong field regime the scalar-tensor theories can behave in a drastically different way. Such a strong non-perturbative effect, called spontaneous scalarization, was discovered numerically  for neutron stars a decade ago \cite{DEF1}. That is why it would be nice and important if we can have exact interior solutions 
giving even only a qualitative picture of the nonlinear structure of a 
static, spherically symmetric star in the presence of the dilaton field. Solving the interior problem in scalar-tensor theories, however, is much difficult and seems to be a hopeless task in the general case due to the presence of the arbitrary coupling function $\alpha(\varphi)$ in the field equations (see eqs. (\ref{EFFE}) below). 
Forced by the mathematical difficulties, we shall focus our attention on a particular 
scalar-tensor theory which allow us to find a sufficiently realistic interior solution 
with the hope that it would give  qualitative insight into the general case of a static, spherically symmetric star. Let us mention that some exact "interior" solutions were previously found 
in Brans-Dicke theory for perfect fluid with equation of state $p=\gamma \rho$ \cite{BK1},\cite{BK2},\cite{K}. Those solutions, however, are singular at the center and have no well defined boundary where the pressure vanishes.

The general form of the extended gravitational action in
scalar-tensor theories is

\begin{eqnarray} \label{JFA}
S = {1\over 16\pi G_{*}} \int d^4x \sqrt{-{\tilde
g}}\left({F(\Phi)\tilde R} - Z(\Phi){\tilde
g}^{\mu\nu}\partial_{\mu}\Phi
\partial_{\nu}\Phi  \right. \nonumber  \\ \left. -2 U(\Phi) \right) +
S_{m}\left[\Psi_{m};{\tilde g}_{\mu\nu}\right] .
\end{eqnarray}

Here, $G_{*}$ is the bare gravitational constant, ${\tilde R}$ is
the Ricci scalar curvature with respect to the space-time metric
${\tilde g}_{\mu\nu}$. The dynamics of the scalar field $\Phi$
depends on the functions $F(\Phi)$, $Z(\Phi)$ and $U(\Phi)$. In
order for the gravitons  to carry positive energy the function
$F(\Phi)$ must be positive. The nonnegativity of the 
dilaton field energy requires that $2F(\Phi)Z(\Phi) +
3[dF(\Phi)/d\Phi]^2 \ge 0$. The action of matter depends on the
material fields $\Psi_{m}$ and the space-time metric ${\tilde
g}_{\mu\nu}$. It should be noted that the stringy generated
scalar-tensor theories, in general, admit a direct interaction
between the matter fields and the dilaton in the Jordan (string)
frame. Here we consider the phenomenological
case when the matter action does not involve the dilaton field in
order for the weak equivalence principle to be satisfied. 

It is much more convenient from a mathematical point of
view to analyze the scalar-tensor theories with respect to the
conformally  related Einstein frame  given by the metric:

\begin{equation}\label {CONF1}
g_{\mu\nu} = F(\Phi){\tilde g}_{\mu\nu} .
\end{equation}

Further, let us introduce the scalar field $\varphi$ (the so
called dilaton) via the equation

\begin{equation}\label {CONF2}
\left(d\varphi \over d\Phi \right)^2 = {3\over
4}\left({d\ln(F(\Phi))\over d\Phi } \right)^2 + {Z(\Phi)\over 2
F(\Phi)}
\end{equation}

 and define

\begin{equation}\label {CONF3}
{\cal A}(\varphi) = F^{-1/2}(\Phi) \,\,\, ,\nonumber \\
2V(\varphi) = U(\Phi)F^{-2}(\Phi) .
\end{equation}

In the Einstein frame the action (\ref{JFA}) takes the form

\begin{eqnarray}
S= {1\over 16\pi G_{*}}\int d^4x \sqrt{-g} \left(R -
2g^{\mu\nu}\partial_{\mu}\varphi \partial_{\nu}\varphi -
4V(\varphi)\right) \nonumber \\ + S_{m}[\Psi_{m}; {\cal
A}^{2}(\varphi)g_{\mu\nu}]
\end{eqnarray}

where $R$ is the Ricci scalar curvature with respect to the
Einstein metric $g_{\mu\nu}$.

Then, the Einstein frame field equations  are

\begin{eqnarray} \label{EFFE}
R_{\mu\nu} - {1\over 2}g_{\mu\nu}R = 8\pi G_{*} T_{\mu\nu}
 + 2\partial_{\mu}\varphi \partial_{\nu}\varphi \nonumber \\  -
g_{\mu\nu}g^{\alpha\beta}\partial_{\alpha}\varphi
\partial_{\beta}\varphi -2V(\varphi)g_{\mu\nu}  \,\,\, ,\nonumber
\end{eqnarray}

\begin{eqnarray}
 \nabla^{\mu}\nabla_{\mu}\varphi = - 4\pi G_{*} \alpha (\varphi)T
+ {dV(\varphi)\over d\varphi} \,\,\, ,
\end{eqnarray}

\begin{eqnarray}
\nabla_{\mu}T^{\mu}_{\nu} = \alpha
(\varphi)T\partial_{\nu}\varphi \,\,\, . \nonumber
 \end{eqnarray}

Here $\alpha(\varphi)= {d\ln({\cal  A}(\varphi))/ d\varphi}$, and
the Einstein frame energy-momentum tensor $T_{\mu\nu}$  is
related to the Jordan frame one ${\tilde T}_{\mu\nu}$ via
$T_{\mu\nu}= {\cal A}^2(\varphi){\tilde T}_{\mu\nu}$. In the case
of a perfect fluid one has

\begin{eqnarray}\label{DPTEJF}
\rho &=&{\cal A}^4(\varphi){\tilde \rho}, \nonumber \\
p&=&{\cal A}^4(\varphi){\tilde p},  \\
u_{\mu}&=& {\cal A}^{-1}(\varphi){\tilde u}_{\mu} . \nonumber
\end{eqnarray}

In what follows we will consider the case $V(\varphi) = 0$.

We have found an exact solution for the following scalar-tensor theory:

\begin{eqnarray}\label{PSTT}
{\cal A}^2(\varphi) = e^{-{2b \varphi\over (2-a)}} \left[(3-a)e^{\varphi\over b} - (2-a)\right]^{2b^2\over (2-a)(3-a)}
\end{eqnarray}

giving the coupling function 

\begin{equation}\label{PSTTCF}
\alpha(\varphi) =b {1- e^{\varphi\over b}  \over (3-a)e^{\varphi\over b} -(2-a) }
\end{equation}

where $a^2 + b^2 =1$ .

In addition to the parameters $a$ and $b$ coming from the particular scalar-tensor theory, the exact solution depends also on two parameters $\mu$ and $R>0$.  In order for our solution to have physical meaning (i.e. positive mass, positive fluid energy density and pressure) the parameters $\mu$ and $a$ must satisfy $\mu a>0$. 

The space-time metric is given by 

\begin{equation}
ds^2 = -e^{2a\lambda}dt^2 + e^{2(1-a)\lambda} {dr^2\over 1- {2\mu G_{*} r^2\over R^3}} + 
e^{2(1-a)\lambda}r^2(d\theta^2 + \sin^2{\theta}d\phi^2 )
\end{equation}

where

\begin{equation}
e^{2\lambda(r)}= {1\over 4} \left[3\left(1-{2\mu G_{*}\over R }\right)^{1/2} - \left(1-{2\mu G_{*} r^2\over R^3} \right)^{1/2} \right]^{2} .
\end{equation}

The dilaton field, the pressure, the fluid energy density and the fluid four-velocity are respectively given by :

\begin{eqnarray}
\varphi(r) = b\lambda(r) - b \ln\left[{3\over2} \left(1-{2\mu G_{*}\over R}\right)^{1/2}\right]  ,
\end{eqnarray} 

\begin{eqnarray}
p(r)={\mu\over {4\pi\over 3}R^3 } { \left(1-{2\mu G_{*} r^2\over R^3} \right)^{1/2} -  \left(1-{2\mu G_{*}\over R }\right)^{1/2}  \over 3 \left(1-{2\mu G_{*}\over R }\right)^{1/2} - \left(1-{2\mu G_{*} r^2\over R^3} \right)^{1/2} } \, e^{-2(1-a)\lambda(r)} ,
\end{eqnarray}

\begin{eqnarray}
\rho(r) = {\mu a\over {4\pi\over 3}R^3} e^{-2(1-a)\lambda(r)}  - 3(1-a) p(r) ,
\end{eqnarray}

\begin{eqnarray}
u= e^{-a\lambda(r)} {\partial \over \partial t } .
\end{eqnarray}

The solution has well defined boundary $r=R$ where, as it can be seen, the pressure vanishes, $p(R)=0$, which determines the star surface. In order for the solution to be 
physically regular everywhere for  $0\le r \le R$, i.e.

\begin{equation}
0< {\cal A}^2[\varphi(r)]<\infty  \,\, \, , 0\le \rho(r) <\infty, \,\,\, 0\le p(r)<\infty, \,\,\,  0<e^{\lambda(r)} < \infty ,
\end{equation}

the parameters  $\mu$ and $R$ must satisfy the inequality

\begin{equation}\label{regcond}
{2|\mu | G_{*}\over R} < {2\over 3} |a| \left(1-{a\over 6}\right) .
\end{equation}

In contrast to the pressure, the fluid energy density does not vanish on the star surface:

\begin{equation}
\rho(R) = {\mu a\over {4\pi\over 3}R^3 }   e^{-2(1-a)\lambda(R)} .
\end{equation}

The metric and the dilaton field of the  interior solution match continuously the external solution

\begin{eqnarray}\label{EXSOL}
ds^2 = - \left(1-{2\mu G_{*}\over r}\right)^{a} dt^2 + \left(1-{2\mu G_{*}\over r}\right)^{-a} dr^2 + \left(1-{2\mu G_{*}\over r}\right)^{1-a} r^2 \left(d\theta^2 + \sin^2\theta d\phi^2 \right), \\
\varphi(r) = {b\over 2}\ln\left(1-{2\mu G_{*}\over r}\right) - b \ln\left[{3\over 2}\left(1-{2\mu G_{*}\over R}\right)^{1/2} \right] .
\end{eqnarray}

The tensor mass (i.e. the  ADM mass in the Einstein frame) can be easily calculated using (\ref{EXSOL}) and is given by $M_{T}= \mu a$. The inequality (\ref{regcond}) then may be written in the form 

\begin{equation}
{2M_{T} G_{*}\over R} < {2\over 3} |a|^2 \left(1-{a\over 6}\right) 
\end{equation}

which can be considered as a (Einstein frame) scalar-tensor version of the Buchdahl inequality.

The asyptotic value of the dilaton field at spatial infinity is 

\begin{equation}
\varphi_{\infty}=\lim_{r\to \infty} \varphi(r) =  - b \ln\left[{3\over 2}\left(1-{2\mu G_{*}\over R}\right)^{1/2} \right]  
\end{equation}

and, obviously, is different from zero. It should be noted that there is a solution 
with zero asymptotic value of the dilaton. Indeed, it is not difficult to see that the field equations (\ref{EFFE}) for the particular scalar-tensor theory defined by (\ref{PSTT}) and (\ref{PSTTCF}) have a solution with a trivial dilaton field, $\varphi=0$ (i.e. pure general relativistic solution).

The Jordan frame solution can be obtained using eqs. (\ref{CONF1}), (\ref{CONF3}) and (\ref{DPTEJF}). 
Since the expressions of the metric, fluid energy density and pressure are rather involved
we will not present them in explicit form.  However,
the qualitative behavior of the fluid pressure, energy density and the gravitational scalar inside the star in the physical Jordan frame can be seen on the figures.

The exact solution presented in this work seems to be the first regular interior solution in scalar-tensor theories of gravity.

I would like to thank V. Rizov for reading the manuscript.    
The present work was partially supported by Sofia University Research Fund.

\newpage

\begin{figure}
\begin{center}
    \includegraphics[width=6.5cm]{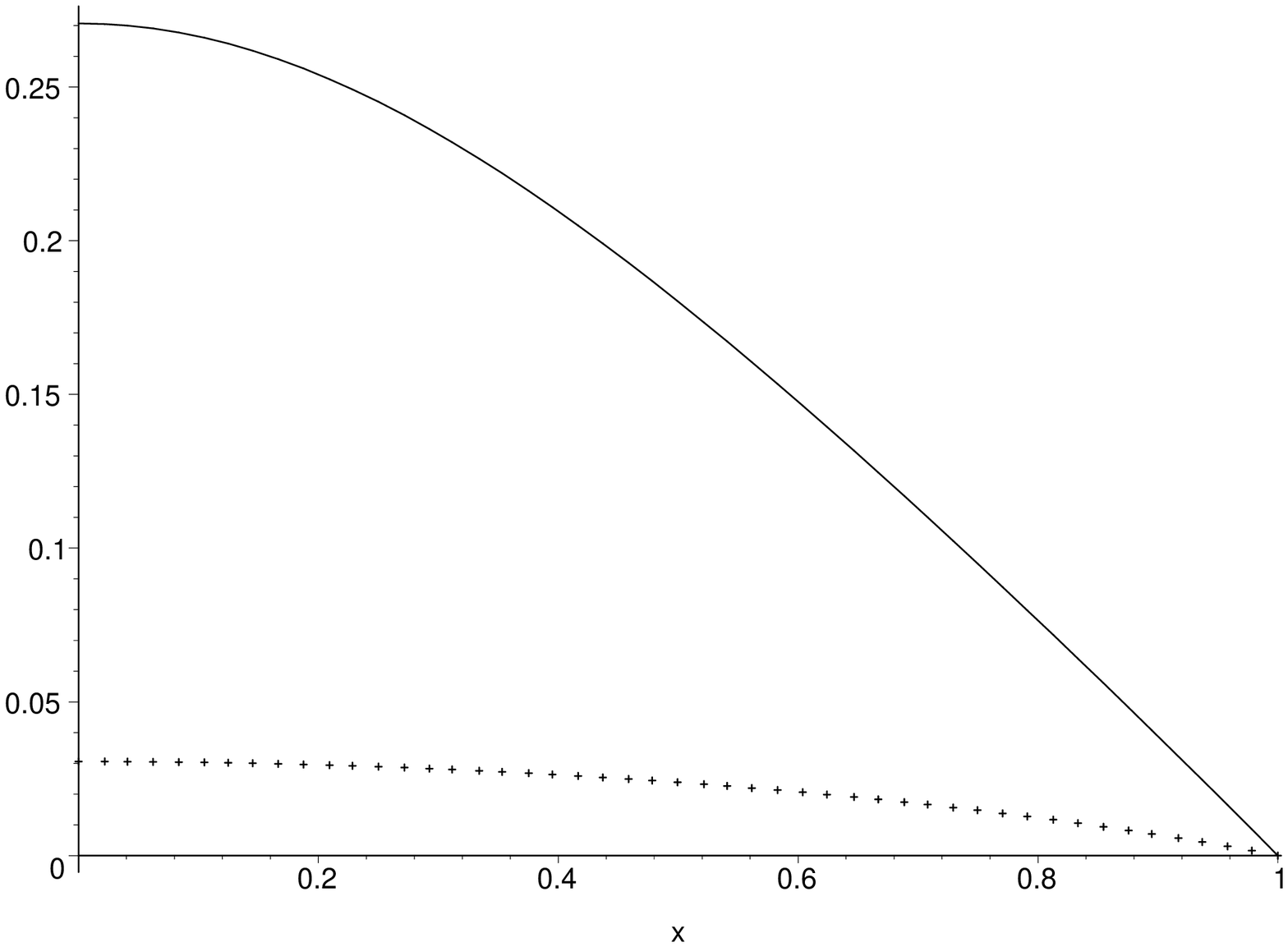}\\
  \caption{Jordan frame pressure ${\tilde p}$ (in units ${|\mu|\over {4\pi\over 3} R^3}$ ) versus the  radial coordinate $x={r\over R}$ for  $|a|=0.9$ and ${2|\mu | G_{*}/R}=0.2$. The solid line represents the case with $a>0$.} \label{fig1}
  \end{center}
\end{figure}

\begin{figure}
\begin{center}
    \includegraphics[width=6.5 cm]{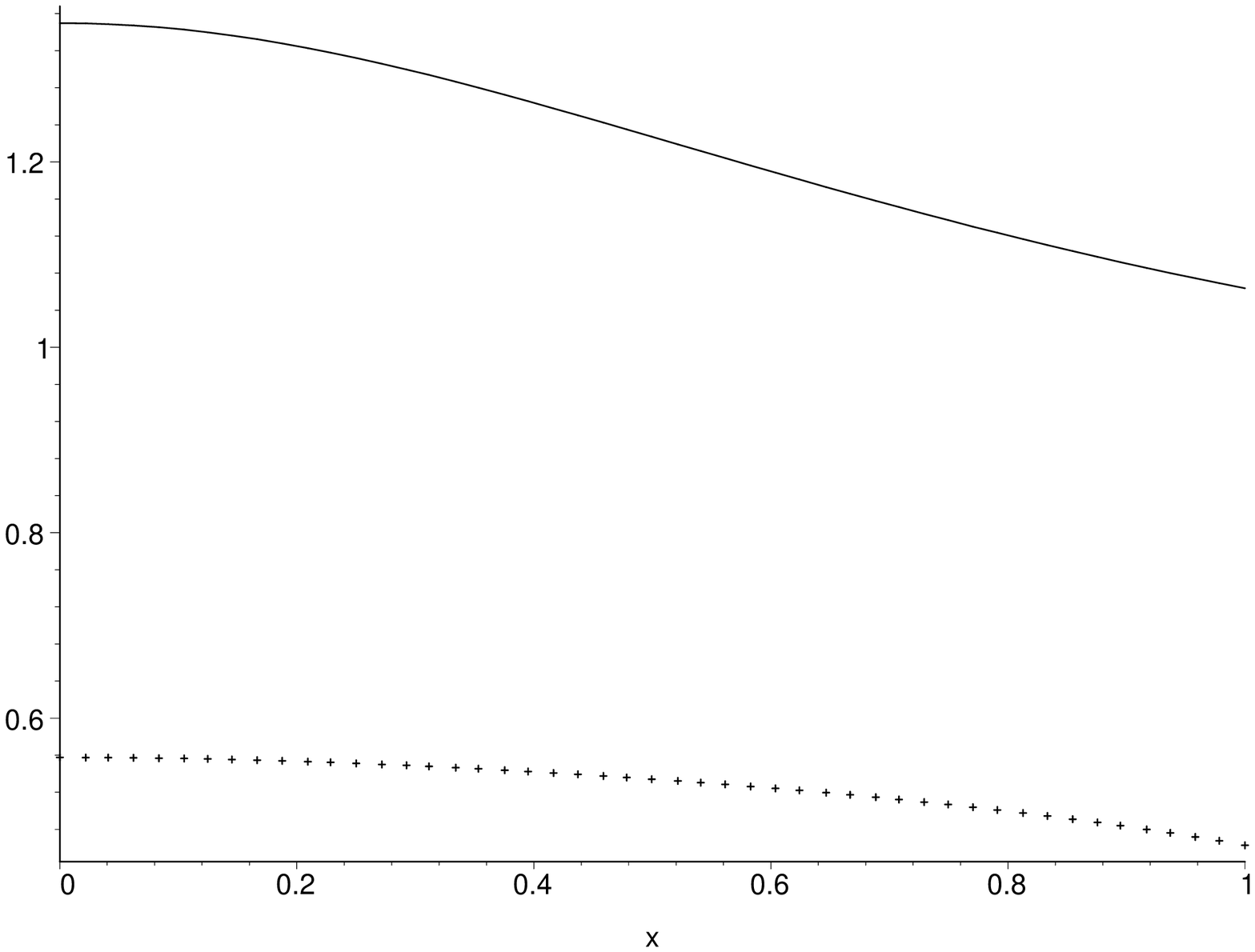}\\
  \caption{Jordan frame energy density ${\tilde \rho}$ (in units ${|\mu| \over{4\pi\over 3} R^3}$ ) versus
  the radial coordinate $x={r\over R}$ for $|a|=0.9$ and $2|\mu | G_{*}/R=0.2$. The solid line represents the case with $a>0$.}\label{fig2}
   \end{center}
\end{figure}

\begin{figure}
\begin{center}
    \includegraphics[width=6.5cm]{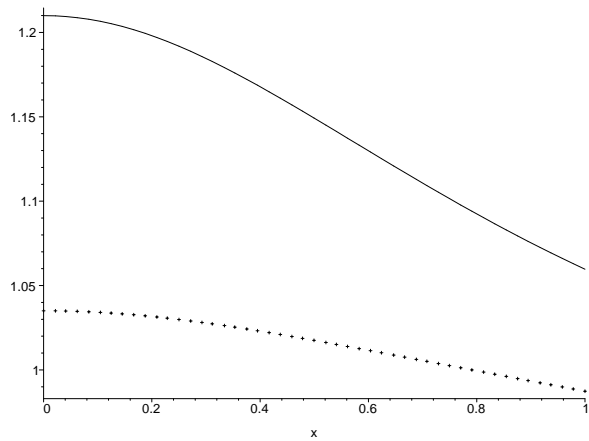}\\
  \caption{The gravitational scalar $F(\Phi)$ as a function  of $x={r\over R}$ for $|a|=0.9$ and $2|\mu | G_{*}/R=0.2$. The solid line represents the case with $a>0$. }\label{fig3}
   \end{center}
\end{figure}

\end{document}